\documentclass[11pt]{article}

\usepackage{a4wide}
\usepackage{epsfig}
\usepackage{graphicx}
\usepackage{multirow}
\usepackage{url}
\usepackage{xspace}

\newcommand{\form}{{\sc Form}\xspace}
\newcommand{\tform}{{\sc TForm}\xspace}
\newcommand{\parform}{{\sc ParForm}\xspace}

\begin{document}

\begin{titlepage}
\noindent

\hfill \begin{minipage}{3.0cm}
Nikhef 2013-036 \hfill \\
TTP13-031       \hfill \\
SFB/CPP-13-80   \hfill \\
October 2013
\end{minipage}
\vspace{8mm}

\begin{center}
\Large
{\bf Code Optimization in FORM} \\
\vspace{1.5cm}
\large
J. Kuipers$^{\, a}$, T. Ueda$^{\, b}$ and J.A.M. Vermaseren$^{\, a}$ \\
\vspace{1.2cm}
\normalsize
{\it $^a$Nikhef Theory Group \\
\vspace{0.1cm}
Science Park 105, 1098 XG Amsterdam, The Netherlands} \\
\vspace{0.5cm}
{\it $^b$Institut f\"ur Theoretische Teilchenphysik,
Karlsruhe Institute of Technology (KIT) \\
\vspace{0.1cm}
D-76128 Karlsruhe, Germany} \\
\vspace{2.0cm}
\large
{\bf Abstract}
\vspace{-0.2cm}
\end{center}
\noindent
We describe the implementation of output code optimization in the open
source computer algebra system \form. This implementation is based on
recently discovered techniques of Monte Carlo tree search to find
efficient multivariate Horner schemes, in combination with other
optimization algorithms, such as common subexpression elimination. For
systems for which no specific knowledge is provided it performs
significantly better than other methods we could compare with. Because the 
method has a number of free parameters, we also show some methods by which 
to tune them to different types of problems.
\vspace{1.0cm}
\end{titlepage}

%


\section{Introduction}
\label{sect::introduction}

One of the uses of computer algebra is to prepare potentially large 
formulas for frequent numerical evaluation. This is particularly the case 
in particle physics. The most widespread way to compute reactions in 
particle physics is by means of perturbative field theory. Even at the one 
loop level (usually the second term in the perturbative expansion) one may 
encouter large numbers of Feynman diagrams, each resulting in a lengthy 
formula\footnote{In special cases other techniques can be used that lead to 
surprisingly simple formulas~\cite{MHV}, but in general these are not 
applicable.}. Such calculations are undertaken to compare theories with 
experimental results. Hence such formulas have to be integrated over the 
region of sensitivity of the detectors that measure these reactions. This 
region is called the experimental acceptance and the only technique that is 
available to integrate over it is Monte Carlo integration. One may have to 
evaluate the formulas millions of times to obtain accurate results. Hence 
it is important to have a representation of the formulas that is as short 
as possible, even if this involves a non-negligible cost during the 
computer algebra phase of the calculation.

Optimizing the output of the formulas can be done in two different ways. 
The first is \emph{domain specific}. This means that specific knowledge 
about the behavior of the formulas is provided to make the formulas 
shorter. An example is an equation in two variables $x$ and $y$, but it is 
known that $x+y$ and $x-y$ are more natural variables and make the formulas 
shorter. For the second way either there is no domain specific knowledge, 
or it is too much work to obtain it. In that case the formula has to be 
treated by generic means. It should be clear that usually the best results 
are obtained when domain specific knowledge is applied first, followed by a 
generic method to clean up what is left.

In computer algebra the challenge is to make a system for the optimization 
of the output of expressions in the absence of domain specific knowledge. 
In addition this system should work reasonably fast, which we interpret as 
subquadratic in the length of the input expression. In the recent past 
several methods have been published in which two of the authors reported on 
new techniques to improve upon existing methods~\cite{kojima,leiserson,
mctshorner}. It turns out that an optimization method based on Monte Carlo 
tree search~\cite{kocsis,chaslot}, a recent search method from artificial 
intelligence and game theory, performs best on the benchmarks that were 
tested. This method has caused much excitement in the field of game theory, 
because it has improved the strength of Go playing computer programs from 
advanced beginner to medium level players, and on small (9x9) boards they 
have reached top level strength. Application of this technique to the field 
of formula simplification has led to sufficiently positive results that we 
have decided to implement it in the computer algebra language 
\form~\cite{form} in such a way that all its users can benefit from it.

Since this is such a new field, we do not yet have extensive
experience with applications and how different types of formulas need
different values for the controlling parameters. Hence we have made an
implementation in which the user has access to these parameters and
can tune them to whole categories of formulas. This means that at the
moment we have a number of default settings which may be changed by
the user. In later versions we may try to have the program tune these
parameters for individual formulas automatically.

The outline of the paper is as follows. In
section~\ref{sect::algorithms} we explain the algorithms that are used
for the simplification. The syntax of the \form implementation is
explained in section~\ref{sect::implementation} including all
parameters that can be set. In section~\ref{sect::examples} we discuss
a number of examples. Section~\ref{sect::effects} is dedicated to
studying the effects of some parameters and the determination of good 
settings for a number of formulas. We finish with remarks about
potential future development. All programs that we use can be obtained
from the \form website at ref.\cite{website}.


\section{Code optimization algorithms}
\label{sect::algorithms}

\subsection{Horner's method}

For optimizing polynomials in a single variable, the textbook
algorithm called \emph{Horner's method} gives an efficient form for
evaluating it~\cite{horner}. It can be written as follows:
\begin{equation}
a(x) = \sum_{i=0}^n a_ix^i = a_0 + x (a_1 + x (a_2 + x(\dots + x\cdot a_n))).
\label{eqn::Horner}
\end{equation}
If the polynomial is of degree $n$ and dense, this form takes $n$
multiplications and $n$ additions to calculate its value.

It is possible to generalize Horner's method for multivariate
polynomials, but this generalization is not unique. First, one of the
variables in the polynomial is selected and Eq.~(\ref{eqn::Horner})
is applied, thereby treating the other variables as constants. Next, a
different variable is chosen and Horner's rule is applied again on
the parts not containing the first variable. This method is repeated
until all variables have been selected. As an example, we consider the
polynomial $a(x,y,z) = y-3x+5xz+2x^2yz-3x^2y^2z+5x^2y^2z^2$ and chose
the variable $x$ first, then $y$ and finally $z$. This results in the
following representation:
\begin{equation}
a(x,y,z) = y+x(-3+5z+x(y(2z+y(z(-3+5z))))).
\label{eqn::Hornerexample}
\end{equation}
This representation takes 8 multiplications and 5 additions to
evaluate, while the original form takes 18 multiplications and 5
additions. This behavior is generic: Horner's method reduces the
number of multiplications and leaves the number of additions constant.

For the multivariate Horner method it is important in which order the
variables are processed. Different orders may lead to huge differences
in the number of operations used to evaluate a
polynomial~\cite{kojima}. Classically, simple greedy algorithms like
sorting the variables by number of occurrences are used to determine
the order~\cite{ceberio}. Recently, two of the authors of this paper 
describedan algorithm based on Monte Carlo tree search to determine more 
efficient orders~\cite{mctshorner}.

\subsubsection{Occurrence order}

In the \emph{occurrence order} all variables are ordered with respect
to their number of occurrences in the polynomial. The variable that
appears most often is the first variable in the
order~\cite{ceberio}. At every step in the multivariate Horner's
method this results in the largest decrease in the number of
operations, because it is the most-occurring variable that is factored
out of the polynomial. This greedy approach usually gives good
results.

Another simple order is the \emph{reverse occurrence order}. As the
name suggests, this order contains the variables sorted with resprect
to the number of occurrences, but with the least-occurring variable
first. This method usually results in Horner schemes that use more
operations than the normal occurrence order to evaluate the
polynomial. However, since the most-occurring variables are now within
the innermost parentheses, more common subexpressions appear, i.e.,
expressions that appear in multiple places in the polynomial. For the
polynomial of Eq.~(\ref{eqn::Hornerexample}) one such common
subexpression is $-3+5z$.

These common subexpressions can be detected by a method called common
subexpression elimination (CSE), see section \ref{sect::CSE}, and be
calculated beforehand. This algorithm of applying a reverse
occurrence order Horner scheme followed by CSE may outperform the
analogous algorithm with the normal occurrence order, if lots of
common subexpressions exist. The difference in performance between
these two algorithms depends primarily on the structure of the input
polynomial.

\subsubsection{Monte Carlo tree search}

\begin{figure}[!ht]
\centering
\includegraphics{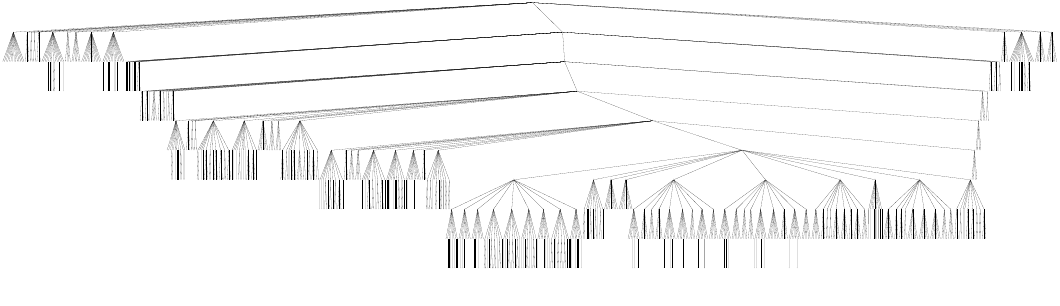}
\caption{The traversed part of an MCTS tree in the search for a good
  Horner scheme.}
\label{fig::MCTStree}
\end{figure}

Recently, the authors of this paper proposed a method to find more
efficient Horner schemes by using Monte Carlo tree search
(MCTS)~\cite{mctshorner}. The different variable orders are
represented by a search tree. The root node indicates that no
variables have been selected. This root node has $n$ children where
$n$ is the number of variables. Traversing down an edge corresponds to
choosing a variable. A node at depth $d$ in the tree represents that
choices are made for the first $d$ variables in the order. Such a
node has $n-d$ children: one for every variable that has not been
selected yet. The MCTS method searches through this tree in an
asymmetric way, where most-promising branches are traversed
first. Fig.~\ref{fig::MCTStree} shows an example of the traversed part
of the MCTS tree after $1\,000$ iterations while looking for an
efficient Horner scheme for a polynomial in 15 variables.

The essence of the MCTS method is the assumption that good solutions
are clustered in branches of the search tree. This means that if a
reasonable solution is found in one of the branches, there is a good
chance that there are more good (and possibly better) solutions in
the same branch. Hence, in the case that a reasonable solution is found it
may pay to search its neighborhood for better solutions. This is
called exploitation. At the same time one should try the untried
branches to see whether they have particularly good or bad solutions,
because the branches tried so far may not have the best solutions. This
process is called exploration. A good MCTS program divides its time
between exploiting previously found favorable branches and exploring
branches about which little is known.

To determine where to look in the search tree for better solutions
several criteria exist. The most-used selection formula is the UCT
(upper confidence level for trees) criterion~\cite{kocsis},
\begin{equation}
UCT_i = \left<x_i\right> + 2 C_p \sqrt{\frac{2\log{n}}{n_i}},
\label{eqn::UCT}
\end{equation}
and the child with the highest UCT value is selected. Here
$\left<x_i\right>$ is the average score of child $i$ over the previous
traversals, $n_i$ is the number of times child $i$ has been visited
before, and $n$ is the number of times the node itself has been
visited. $C_p$ is a problem-dependent constant that should be
determined empirically.

If one enters a node that has never been visited before, the remaining 
decisions are all random. This means that one can come to a solution very 
fast after which one can evaluate the quality of this solution. In the game 
of Go, for example, this idea of selecting random moves until the game is 
over and the outcome can the determined might seem like complete madness at 
first. On the other hand: if there is a good move, one expects that the 
average score in the branch with this move will be higher than in the other 
branches and it pays to look a bit better in this branch. Of course there 
is the risk that a branch with a very good move is missed because an 
earlier random continuation was particularly bad. Hence one should never 
exclude branches completely and this is the role of the second term in 
Eq.~(\ref{eqn::UCT}); eventually the branch will be selected again.

A complete description of MCTS in general and its applications can be
found in Ref.~\cite{browne}, while a more detailed description of MCTS
for Horner schemes including pseudocode is in Ref.~\cite{mctshorner}.

As mentioned above, for MCTS to work we need clustering of good and
bad solutions. Hence the selection of the tree structure is dominantly
important. In the case of Horner schemes the choice of whether the
outermost variable or the innermost is selected first can make all the
difference, in the same way as using occurrence order or reverse
occurrence order does. We found that this choice depends on the
formula to be optimized. A clear example of this is given in
section~\ref{sect::effects}.

\subsection{Common subexpression elimination}
\label{sect::CSE}

Large expressions may contain many common subexpressions, in other
words, subexpressions that appear in the equation multiple times. A
small example of this is the subexpression $-3+5z$ in
Eq.~(\ref{eqn::Hornerexample}). These can be replaced by a temporary
variable to reduce the number of operations, as the following code
shows:
\begin{equation}
\begin{array}{l}
Z_1 = -3+5z \\
a = y+x(Z_1+x(y(2z+y(zZ_1)))).
\end{array}
\end{equation}
This code uses only 7 multiplications and 4 additions, while the
original expression after applying Horner's method uses 8
multiplications and 5 additions.

\begin{figure}[b!]
\centering
\begin{tabular}{cc}
\includegraphics{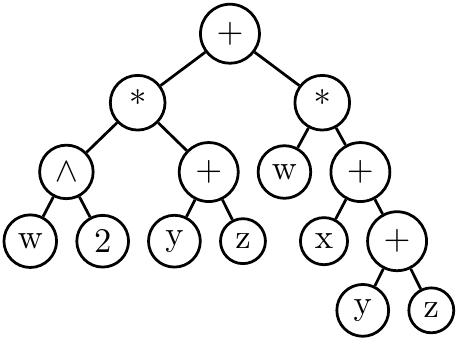}
&
\includegraphics{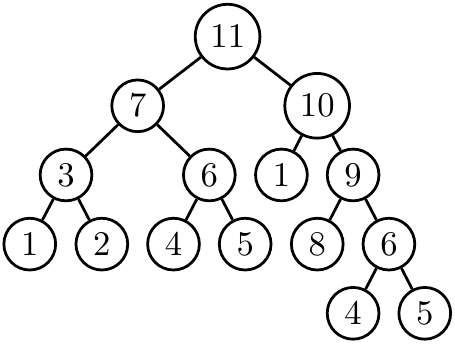}
\end{tabular}
\caption{A binary expression tree for the polynomial $w^2y + w^2z + wx
  + wy + wz$ and the labeling of the nodes after CSE.}
\label{fig::CSE}
\end{figure}

Common subexpressions can be detected and eliminated by the method of
\emph{common subexpression elimination} (CSE) in time linear in the
size of the input expression~\cite{compilers}. First, the expression
is represented as a binary tree with the operators as inner nodes and
the variables and numbers as leaves. The two children of an operator
denote its left and right operand. The left panel of
Fig.~\ref{fig::CSE} shows an example of a binary expression tree for
the polynomial $w^2y + w^2z + wx + wy + wz$.

This expression tree is traversed and each subtree is assigned a
number starting with the leaves. Equal subtrees are assigned equal
numbers, since these represent common subexpressions. An associative
array (e.g., a hash table or balanced binary tree)~\cite{cormen} is
used to map subexpressions (i.e., a number, a variable or an operator
combined with the identifiers of its two operands) to the
identifiers. When moving up in the tree, this map facilitates checking
fast whether a subtree is encountered before or should be assigned a
new identifier. The right panel of Fig.~\ref{fig::CSE} shows the
labeling of the subtrees after CSE.

This CSE method works particularly well in combination with Horner's
method if the variables that appear often in common subexpressions are
chosen as innermost variables in the Horner order. In this way, many
common subexpressions appear and the cost of evaluation is greatly
reduced. 

\begin{figure}[!ht]
\centering
\begin{tabular}{cc}
\includegraphics{tree1}
&
\includegraphics{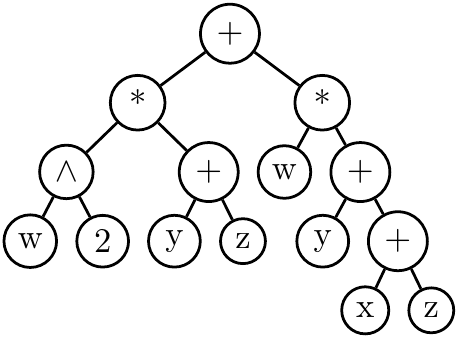}
\end{tabular}
\caption{Two different binary expression trees for representing the
  polynomial $w^2y + w^2z + wx + wy + wz$. Only in the left one, CSE
  detects the common subexpression $y+z$.}
\label{fig::multipletrees}
\end{figure}

However, this method can also miss many common subexpressions. We
consider again the polynomial $w^2y + w^2z + wx + wy + wz$, which can
be represented as a binary expression tree in multiple ways of which
two of them are shown in Fig.~\ref{fig::multipletrees}. When using
CSE, the common subexpression $y+z$ is missed in the right
representation, but is detected in the left one. Unfortunately, it
seems impossible to detect such structures in time linear in the size
of the input expression.

\subsection{Greedy optimizations}
\label{sect::greedy}

\begin{figure}[!ht]
\centering 
\begin{tabular}{cc}
\includegraphics{tree2}
&
\includegraphics{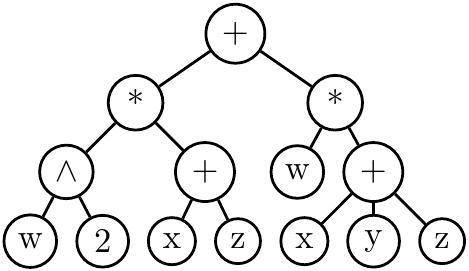}
\end{tabular}
\caption{An expression tree of $w^2y + w^2z + wx + wy + wz$ before
  (left) and after (right) merging operators.}
\label{fig::mergedtree}
\end{figure}

CSE is not able to detect all forms of common subexpressions as is
illustrated by Fig.~\ref{fig::multipletrees}. Therefore another method
is needed to detect them and reduce the evaluation cost even
further. To do so, equal operators are merged first. Two nodes in the
expression tree are merged if they contain equal operators and they
are each other's parent and child. The new node has as children all
children of merged nodes. Note that the expression tree is not a
binary tree anymore after operators are merged. This merging process
is shown in Fig.~\ref{fig::mergedtree}.

After merging equal operators the evaluation code is generated. The
lines of this code are called \emph{intermediate expressions}. We
demand that these intermediate expressions contain only operators of
one single type, i.e., either additions or multiplications. Producing
code of this form has already been suggested in Ref.~\cite{breuer} for
applying Breuer's growth algorithm. The resulting code for the merged
tree of Fig.~\ref{fig::mergedtree} is the left hand side of
Eq.~(\ref{eqn::greedycode}).

\begin{equation}
\begin{array}{lcl}
Z_1 = w^2       && Z_1 = w^2 \\
Z_2 = y+z       && Z_2 = y+z \\
Z_3 = Z_1 * Z_2 & \multirow{2}{*}{$\longrightarrow$} & Z_3 = Z_1 * Z_2 \\
Z_4 = x+y+z     && Z_4 = x+Z_2 \\
Z_5 = w*Z_4     && Z_5 = w*Z_4 \\
a = Z_3 + Z_5   && a = Z_3 + Z_5
\end{array}
\label{eqn::greedycode}
\end{equation}

After writing the code in this way, every expression is scanned
through and the number of occurrences of certain small subexpressions
is counted. The subexpressions counted are of the form
\begin{equation}
x^n,\quad x\cdot y,\quad c\cdot x,\quad x+c,\quad x+y\quad
\mbox{or}\quad x-y.
\end{equation}
In these small subexpressions $x$ and $y$ are either variables of the
polynomial ($w,x,y,z$ in Eq.~(\ref{eqn::greedycode})) or intermediate
variables ($Z_i$ in Eq.~(\ref{eqn::greedycode})), and $c$ is a
coefficient. For each intermediate expression we loop over all pairs
of terms and count the corresponding subexpressions. This takes time
proportional to the sum of the squares of the lengths of the
intermediate expressions, and is therefore much slower than CSE.

Next, the subexpressions that occur multiple times are determined,
because replacing them by new intermediate expressions reduces the
total number of operations of the evaluation code. A fraction of these
optimizations, that give the largest decrease in evaluation cost, is
performed and subsequently a new list of optimizations is generated.
This process is repeated until no more optimizations are found.

This algorithm is different from Breuer's growth
algorithm~\cite{breuer,hulzen}, where a larger subexpression that generates a
bigger decrease in cost is determined. Finding that subexpression is
computationally harder than counting our small subexpressions, and
this has to be calculated repeatly. We believe the two methods are
more or less equivalent though, since the larger substitution may be
viewed as a chain of small ones, but have no proof of this.

\subsection{Partial factorization}

The algorithm of greedy optimizations is good for detecting a lot of
common subexpressions. It does not find optimizations of the following
form though:
\begin{equation}
\begin{array}{lcl}
                    && Z_1 = y*z \\
Z_1 = x*y*z         && Z_2 = z^2 \\
Z_2 = x*z^2         & \longrightarrow & Z_3 = 2+Z_1+Z_2 \\
a   = 2*x+y+Z_1+Z_2 && Z_4 = x*Z_3 \\
                    && a   = y+Z_4
\end{array}
\label{eqn::partialfact}
\end{equation}
Here, the variable $x$ is factored out of a number of terms, thereby
reducing the number of operations. These kind of optimizations are
done by the \emph{partial factorization} method. For each intermediate
expression and the intermediate expressions in its operands, the
number of occurrences of each variable is counted. If a variable
occurs two or more times, the code is optimized analogous to
Eq.~(\ref{eqn::partialfact}). In our code optimization algorithms,
this method of partial factorization is intertwined with the greedy
optimizations, so that both methods optimize the evaluation code in
turns. This generally gives good resulting code.

This partial factorization method has some similarities with the
hypergraph method that performs syntactic
factorizations~\cite{leiserson}. In this method a polynomial $a$ is 
factorized as $a=b\cdot c$, such that the number of terms in $a$ is
the product of the numbers of terms in $b$ and $c$. The factors $b$
and $c$ have to be more or less independent for this to happen. In
this case the substitution rule of Eq.~(\ref{eqn::partialfact}) often
finds the corresponding factorization after a few iterations
intertwined with the greedy optimizations. This is shown by the
sequence underneath.
\begin{equation}
\begin{array}{lclclcl}
Z_1 = w*y             && Z_5 = y+z    && Z_5 = y+z    && \\
Z_2 = w*z             && Z_6 = w*Z_5  && Z_6 = w*Z_5  && Z_5 = y+z   \\
Z_3 = x*y             &\longrightarrow& Z_7 = y+z    &\longrightarrow& Z_8 = x*Z_5  &\longrightarrow& Z_9 = w+x  \\
Z_4 = x*z             && Z_8 = x*Z_7  && a   = Z_6+Z_8 &&  a = Z_5*Z_9 \\ 
a   = Z_1+Z_2+Z_3+Z_4 && a   = Z_6+Z_8 && && 
\end{array}
\end{equation}

\subsection{Recycling variables}

The algorithms to reduce the number of operations (Horner schemes
followed by CSE, greedy optimizations and/or partial factorizations)
usually result in evaluation code that contains a huge number of
intermediate variables, namely one per intermediate expression. Many
of them are only used in a small part of the code and can be discarded
afterwards. Reusing these variables often greatly reduces the number
of temporary variables, which leads to more efficient code and faster
compile times when compiling it with traditional compilers.

To generate the final code and recycle the variables, the intermediate
expressions are sorted in \emph{depth-first
  order}~\cite{compilers}. The directed acyclic graph of intermediate
expressions is traversed with a depth-first search and an expression
is added to the code after all its children are traversed. Next, the
life ranges of all expressions are determined by looking at their
first and last appearances. Subsequently, new expressions are
renumbered to the lowest available number at that stage. This method
of renumbering variables is known as \emph{linear scan register
  allocation}~\cite{poletto} and works much faster for large numbers
of variables than the traditional graph coloring algorithms used in
many compilers. After appying this final step, the code of
Eq.~(\ref{eqn::greedycode}) becomes the following
\begin{equation}
\label{eqn::greedyoutcome}
\begin{array}{lcl}
Z_1 = w^2     && Z_1 = w^2 \\
Z_2 = y+z     && Z_2 = y+z\\
Z_3 = Z_1 * Z_2 & \multirow{2}{*}{$\longrightarrow$} & Z_1 = Z_1 * Z_2 \\
Z_4 = x+Z_2   && Z_2 = x+Z_2 \\
Z_5 = w*Z_4   && Z_2 = w*Z_2 \\
a = Z_3 + Z_5 && a = Z_1 + Z_2
\end{array}
\end{equation}
and only two temporary variables are used to evaluate the polynomial.


\section{FORM implementation}
\label{sect::implementation}

\subsection{Calling optimization routines}

There are several statements that concern the optimization of an expression 
and its writing. New options of the \verb|Format| statement determine 
whether optimization is applied and how this is done. Regular output can be 
printed with the \verb|Print| statement. If the format is set correctly the 
output is optimized. Such output does not affect the stored version of the 
expressions that are optimized. It is only the output that takes this 
representation.

This output, however, is usually not what the user intends. Things are 
different when the \verb|#optimize| instruction is used. This optimizes a 
single expression and replaces the original formula by the new one and 
keeps all temporary statements like the ones in the r.h.s. of 
Eq.~(\ref{eqn::greedyoutcome}) in memory. This can only be done for one 
expression at a time. As soon as a second expression is optimized the 
results of the previous optimization are lost and so is the entire previous 
expression. The reason for this behavior is explained below. The optimized 
expression and its supportive statements can be written to the standard 
output or a file with the \verb|#write| instruction. This last method is 
very powerful and allows the automatic construction of complete programs.

The \verb|#write| instruction has in the format string the new
specifier \verb|%O| which writes the intermediate statements that are
prepared by the most recent \verb|#optimize| instruction. It uses the 
current format settings (as for fortran or C) and the current settings for 
the extra symbols. Additionally, when for instance \verb|%5O| is specified, 
all lines in the output will have 5 spaces at the start of the line. The 
format specifier \verb|%E| can write the optimized expression.

To declare the array for the temporary variables, the user can access 
the maximum and minimum range of the array used in the optimized code by 
the preprocessor variables {\tt optimmaxvar\_} and {\tt optimminvar\_}, 
respectively.

Finally, there is a \verb|#clearoptimize| instruction that clears the 
optimization buffer and removes the optimized expression.

If an expression contains brackets, the outsides of the brackets are not 
treated in the optimizations. This can be used to create simultaneous 
optimizations as in the second example in the next section.

\subsection{Interaction with extra symbols}

Some operations in \form have been implemented only for multivariate 
polynomial expressions and the same may hold for some procedures that have 
been constructed by users. To facilitate this \form version 4.0 has been 
equipped with the statement \verb|ToPolynomial| that replaces all objects 
(like functions or dotproducts) by system defined symbols, called extra 
symbols. It is also possible to undo this substitution at a later stage 
with the \verb|FromPolynomial| statement. The \verb|ExtraSymbol| statement 
controls the output representation of the extra symbols in the case they 
have to be printed.

The implementation of code optimization needs the notion of systems defined 
intermediate variables. We define these temporary variables as extra 
symbols in addition to the ones that may exist already.

The definitions of the extra symbols that are introduced by the 
\verb|ToPolynomial| statement are unique as in
\begin{verbatim}
    Z1_ = g(x)
    Z2_ = g(y)
    Z3_ = f(x)
\end{verbatim}
and hence we can store them in rather simple and semi permanent buffers. 

The intermediate variables introduced during optimization can be reused as we 
explained before and hence the definitions of the corresponding extra 
symbols are not unique as in
\begin{verbatim}
    Z1_=x + y;
    Z2_=y*Z1_;
    Z1_=Z1_ + z;
    Z1_=z*Z1_;
    Z3_=x^2;
    Z1_=Z1_ + Z3_ + Z2_;
    F=2*Z1_;
\end{verbatim}
where the definitions of \verb:Z1_: and \verb:Z2_: are overwritten (in the 
case of \verb:Z1_: even more than once). This, in combination with the 
potential number of generated statements is the reason that this category 
of extra symbols is stored in a different, more temporary way.
We allocate a special buffer for them and because of its potential size, we 
do not want to allocate more than a single buffer. If there would be more 
than a single one, there would be additional complications concerning the 
numbering of the variables in the different buffers: are they independent, 
causing several instances of \verb:Z1_:, or should they be numbered 
consecutively? We have opted for a single buffer, but with a mechanism for 
optimization of more than one object at the same time.

\subsection{Optimization options of the Format statement}

The \verb|Format| statement has a number of options to control the
code optimization. The easiest to use are the following:

\begin{description}
\item[O0] Switches off all optimizations and prints the output the
  normal \form way. This is the default.

\item[O1] Activates the lowest level of optimization. It is very fast,
  i.e., linear in the size of the expression, and gives reasonably
  efficient code.

\item[O2] Activates the medium level of optimization. This is slower
  than the previous setting, but usually gives better results.

\item[O3] Activates the highest level of optimization. It can be
  rather slow, but usually gives even better results.
\end{description}

These levels of optimization refer to some default settings of all
controlling parameters. These default values are in
Tab.~\ref{tbl:defaults}. It is also possible to set each parameter
individually to fine-tune the optimization process. The parameters
that can be set are divided in several categories. First, it is
possible to set which Horner schemes are tried:

\begin{description}
\item[Horner=(Occurrence $|$ MCTS)] Determines whether a (possibly
  reverse) occurrence order Horner scheme is used or whether MCTS is
  employed to find Horner schemes.

\item[HornerDirection=(Forward $|$ Backward $|$ ForwardOrBackward $|$] \hfill
  {\bf ForwardAndBackward)}
  For the occurrence order, forward selects the
  normal one and backward selects the reverse one. ForwardOrBackward
  and ForwardAndBackward try both. For MCTS, forward starts selecting
  the first variables in the Horner scheme and backward starts with
  the last ones. ForwardOrBackward tries both of these
  schemes. ForwardAndBackward fill the order from both sides
  simultaneously, resulting in more options, but also a much larger
  search tree.
  
\end{description}

In the case of MCTS there are various parameters that can control the
search process:

\begin{description}
\item[MCTSConstant=$<$\emph{value}$>$] This is the constant $C_p$ in
  Eq.~(\ref{eqn::UCT}).
\item[MCTSNumExpand=$<$\emph{value}$>$] The number of times the tree
  is traversed and hence the number of times that a Horner scheme is
  constructed.
\item[MCTSNumKeep=$<$\emph{value}$>$] The number of best solutions
  that will be remembered.
\item[MCTSNumRepeat=$<$\emph{value}$>$] As we will see in the
  section~\ref{sect::effects} sometimes it is more advantageous to run
  a new tree search several times, each with a smaller number of
  expansions. This parameter tells how many times we will run with a
  new tree. The total number of tree traversals is the product of 
  MCTSNumRepeat and MCTSNumExpand.
\item[MCTSTimeLimit=$<$\emph{value}$>$] The maximum time in seconds
  that is used when searching through the tree.
\end{description}

The Horner methods generate a number of Horner schemes: one or two in
the case of occurrence order schemes, depending of the direction
parameter, and a number equal to MCTSNumKeep in the case of
MCTS. Next, for each stored Horner scheme other optimizations are
performed as determined by the following parameter:

\begin{description}
\item[Method=(None $|$ CSE $|$ Greedy $|$ CSEGreedy)] Determines what
  method is used for optimizing the generated Horner schemes. CSE
  performs common subexpression elimination (see
  section~\ref{sect::CSE}) and Greedy performs greedy optimizations
  (see section~\ref{sect::greedy}). CSEGreedy performs CSE followed by
  greedy optimizations; usually this is somewhat faster than just
  greedy optimizations, but it gives slightly worse results. The
  option None does nothing after applying the Horner scheme and is
  only useful for debugging purposes.
\end{description}

When the method of greedy optimizations is used, repeatedly all
optimizations are determined and a few of them are performed. The
following parameters are used to tune the greedy method:
\begin{description}
\item[GreedyMaxPerc] The percentage of the possible optimizations that is
  performed.
\item[GreedyMinNum] The minimum number of possible optimizations that
  is performed.
\item[GreedyTimeLimit] The maximum time in seconds that is spent in
  the process of greedy optimization.
\end{description}
Experimentation shows that the influence of the first two parameters is not 
very big. They might however be useful when the greedy method is expanded 
with the recognition of more complicated substructures.

Additionally, there are two more general settings:
\begin{description}
\item[Stats=(On $|$ Off)] This parameter determines whether statistics
  of the optimization are shown.
\item[TimeLimit=$<$\emph{value}$>$] This set both the MCTSTimeLimit
  and the GreedyTimeLimit to half of the given value.
\end{description}

Finally there are a few options that can help very much with debugging:
\begin{description}
\item[DebugFlag=(On $|$ Off)] 
In the case that the value is On, the list of temporary variables is 
printed in reverse order with the the string "id " in front. This makes 
them into a set of \form substitutions that undo the optimizations. One 
can use this for instance to make sure that the optimized code is identical 
to the original.
\item[PrintScheme=(On $|$ Off)]
This option (when On) will print the Horner scheme. That is the order in 
which the variables were taken outside parentheses.
\item[Scheme=(list of symbols)] The list should be enclosed by parentheses 
and the symbols should be separated by either blanks or comma's. This 
option will fix the Horner scheme to be used.
\end{description}

{ \small
\begin{table}[!ht]
\centering
\begin{tabular}{|l|c|c|c|}
\hline
                &     O1      &    O2       & O3 (default) \\
 \hline
Horner          &  occurrence &  occurrence &   MCTS   \\
HornerDirection &     OR      &      OR     &    OR    \\
MCTSConstant    &    ---      &     ---     &   1.0    \\
MCTSNumExpand   &    ---      &     ---     &  1000    \\
MCTSNumKeep     &    ---      &     ---     &   10     \\
MCTSNumRepeat   &    ---      &     ---     &    1     \\
MCTSTimeLimit   &    ---      &     ---     &    0     \\
Method          &    cse      &    greedy   &  greedy  \\
GreedyMinNum    &    ---      &     10      &   10     \\
GreedyMaxPerc   &    ---      &      5      &    5     \\
GreedyTimeLimit &    ---      &      0      &    0     \\
Stats           &    off      &     off     &   off    \\ 
TimeLimit       &     0       &      0      &    0     \\
\hline
\end{tabular}
\caption{Values for the various parameters in the predefined
  optimization levels. OR stands for ForwardOrBackward.}
\label{tbl:defaults}
\end{table}
} 

All options should be specified in a single format statement and be
separated either by commas or blank spaces. When
\verb|Format Optimize| is used, first the default settings are taken
and then the options that are specified overwrite them. It is allowed
to have the O1, O2, O3 optimization specifications followed by
options. In that case the program first sets the values of those
specifications and then modifies according to what it encounters in
the rest of the statement.


\section{Examples}
\label{sect::examples}

\subsection{Optimizing a single expression}

The first example shows a rather simple use.
\begin{verbatim}
    Symbols x,y,z;
    Off Statistics;
    Local F = 6*y*z^2+3*y^3-3*x*z^2+6*x*y*z-3*x^2*z+6*x^2*y;
    Format O1,stats=on;
    Print;
    .end
      Z1_=y*z;
      Z2_= - z + 2*y;
      Z2_=x*Z2_;
      Z3_=z^2;
      Z1_=Z2_ - Z3_ + 2*Z1_;
      Z1_=x*Z1_;
      Z2_=y^2;
      Z2_=2*Z3_ + Z2_;
      Z2_=y*Z2_;
      Z1_=Z2_ + Z1_;
      F=3*Z1_;
*** STATS: original  1P 16M 5A : 23
*** STATS: optimized 0P 10M 5A : 15
\end{verbatim}
The statistics show that we started with 23 operations (one power which 
counted double because it was a third power, 16 multiplications and 5 
additions) and that we are left with 15 operations. Note that squares are 
counted like a single multiplication. If we run this program 
with the option O2 we obtain
\begin{verbatim}
      Z1_=z^2;
      Z2_=2*y;
      Z3_=z*Z2_;
      Z2_= - z + Z2_;
      Z2_=x*Z2_;
      Z2_=Z2_ - Z1_ + Z3_;
      Z2_=x*Z2_;
      Z3_=y^2;
      Z1_=2*Z1_ + Z3_;
      Z1_=y*Z1_;
      Z1_=Z1_ + Z2_;
      F=3*Z1_;
*** STATS: original  1P 16M 5A : 23
*** STATS: optimized 0P 9M 5A : 14
\end{verbatim}
and with O3 we have
\begin{verbatim}
      Z1_=x + z;
      Z2_=2*y;
      Z3_=Z2_ - x;
      Z1_=z*Z3_*Z1_;
      Z3_=y^3;
      Z2_=x^2*Z2_;
      Z1_=Z1_ + Z3_ + Z2_;
      F=3*Z1_;
*** STATS: original  1P 16M 5A : 23
*** STATS: optimized 1P 6M 4A : 12
\end{verbatim}
It is possible to obtain an even better decomposition, but this requires 
simplifications of the type $x^2+xz+z^2\rightarrow (x+z)^2-xz$ which is not 
within the scope of the simplifications we apply. Similarly $x^2+2xz+z^2$ 
will not be seen as a square. Such simplifications require entirely 
different algorithms which should be executed before the \form algorithms 
are applied.


\subsection{Simultaneous optimization}

Imagine we have to compute two objects F and G. If we can compute the 
common subexpressions of both only once we may save much. To do this we put 
the two expressions into a single expression H in such a way that we can 
separate them again by bracketting in the extra variable u. If the 
expression is bracketted in terms of u at the moment of optimization \form 
will know that it is not allowed to combine terms like \verb:u*a+u^2*a: 
into \verb:(u+u^2)*a:. The output expression in H still contains the 
variable u and when we bracket again in u, we can recover the individual 
expressions, but now in their optimized version.
\begin{verbatim}
    Symbols x,y,z,u;
    ExtraSymbols,array,tmp;
    Off Statistics;
    Local F = (x+y+z)^2;
    Local G = (x+2*y+z)^2;
    .sort
    Format O3;
    Local H=u*F+u^2*G;
    B	u;
    .sort
    #optimize H
    B	u;
    .sort
    Local F1 = H[u];
    Local G1 = H[u^2];
    .sort
    #write <> "%5O"

     tmp(1)=z + 2*x;
     tmp(1)=tmp(1)*z;
     tmp(2)=x^2;
     tmp(1)=tmp(1) + tmp(2);
     tmp(2)=z + x;
     tmp(3)=2*tmp(2) + y;
     tmp(3)=y*tmp(3);
     tmp(3)=tmp(3) + tmp(1);
     tmp(2)=y + tmp(2);
     tmp(2)=y*tmp(2);
     tmp(1)=4*tmp(2) + tmp(1);

    #write <> "\n     F=%e     G=%e",F1,G1

     F=tmp(3);
     G=tmp(1);

    .end
\end{verbatim}
The same program, but now without writing the output and showing 
statistics, also for the optimization of the individual expressions, shows 
the gain by doing the optimization simultaneously.
\begin{verbatim}
    Symbols x,y,z,u;
    ExtraSymbols,array,tmp;
    Off Statistics;
    *Format nospaces;
    Local F = (x+y+z)^2;
    Local G = (x+2*y+z)^2;
    .sort
    Local H=u*F+u^2*G;
    B	u;
    *Print;
    .sort
    Format O3,stats=on;
    #optimize H
*** STATS: original  0P 19M 10A : 29
*** STATS: optimized 0P 7M 7A : 14
    .sort
    #optimize F
*** STATS: original  0P 9M 5A : 14
*** STATS: optimized 0P 5M 5A : 10
    #optimize G
*** STATS: original  0P 10M 5A : 15
*** STATS: optimized 0P 5M 5A : 10
    .end
\end{verbatim}
The counting of the number of operations in the original version of H takes 
into account that the brackets fulfil a special role here.

Of course, by using factorization one can write the above expressions 
shorter than this, but that is not the point here. In general we do not 
apply factorization because that succeeds only in very rare cases and 
it can be very slow on large expressions. Potential factorization is 
considered `domain specific' and falls under the responsibility of the 
user.

\subsection{Optimizing resultants}

The third example is bigger. It comes from ref.~\cite{leiserson}. We compute 
the resultant of two polynomials $A = \sum_{i = 0}^m a_i x^i$ and $B = 
\sum_{i = 0}^n b_i x^i$. This is a $(m+n)\times(m+n)$ determinant and gives 
a polynomial in $m+n+2$ variables. The program is rather short and can go 
to rather large values of $m$ and $n$. The complete program can be picked 
up from the \form website (see ref~\cite{website}). Here we show the final 
part of the program in which F is a stored expression containing the 
$7\times 5$ resultant:
\begin{verbatim}
   Format O1,stats=on;
   L	F1 = F;
   .sort
   #message CPU time till now: `time_' sec.
   #optimize F1
   .store
   Format O2,stats=on;
   L	F2 = F;
   .sort
   #message CPU time till now: `time_' sec.
   #optimize F2
   .store
   Format O3,stats=on,mctsconstant=0.1;
   L	F3 = F;
   .sort
   #message CPU time till now: `time_' sec.
   #optimize F3
   .store
   #message CPU time till now: `time_' sec.
   .end
\end{verbatim}
and the output is
\begin{verbatim}
   ~~~CPU time till now: 0.34 sec.
   *** STATS: original  12044P 106580M 11379A : 142711
   *** STATS: optimized 41P 11745M 8359A : 20210
   ~~~CPU time till now: 0.75 sec.
   *** STATS: original  12044P 106580M 11379A : 142711
   *** STATS: optimized 42P 8873M 7417A : 16398
   ~~~CPU time till now: 6.12 sec.
   *** STATS: original  12044P 106580M 11379A : 142711
   *** STATS: optimized 25P 5551M 5561A : 11171
   ~~~CPU time till now: 189.73 sec.
     189.73 sec out of 189.79 sec
\end{verbatim}
The times are on an Opteron 2.6 GHz processor. The first time is the time 
needed to obtain the $12\times 12$ determinant. The times are cumulative. 
It is possible to obtain even better values at the cost of more CPU time. 
It is also possible to run the program with \tform or \parform to 
get the benefit of parallelization. For the O1 and O2 optimizations this 
gives a modest improvement in the time, because it runs the scheme twice: 
once with the Horner scheme in forward mode and once in backward mode (as 
in the forwardorbackward setting). These can be executed in parallel. The 
Monte Carlo approach of the O3 level is very suitable for parallelization. 
This is illustrated by the following \tform run with 4 workers:
\begin{verbatim}
   ~~~CPU time till now: 0.86 sec.
   *** STATS: original  12044P 106580M 11379A : 142711
   *** STATS: optimized 41P 11745M 8359A : 20210
   ~~~CPU time till now: 1.47 sec.
   *** STATS: original  12044P 106580M 11379A : 142711
   *** STATS: optimized 42P 8873M 7417A : 16398
   ~~~CPU time till now: 7.17 sec.
   *** STATS: original  12044P 106580M 11379A : 142711
   *** STATS: optimized 25P 5306M 5364A : 10730
   ~~~CPU time till now: 220.27 sec.
     0.37 sec + 220.00 sec: 220.38 sec out of 58.81 sec
\end{verbatim}
The times printed during the running are the times of the master processor 
and hence rather meaningless. But the final time involves all workers and 
the 58.81 sec is the real time that passed. The difference between the 
total CPU time of the sequential run (189.79 sec) and the run with 4 
workers (220.38 sec) is partially due to overhead and partially to 
bus-congestion. Yet it is clear that we gain more than a factor 3. The 
different number of operations in the O3 optimization is due to the Monte 
Carlo nature and the fact that each worker has its own initialization of 
the random number generator. In general such multicore runs are not 
deterministic anyway, because the order in which objects are processed is 
not fixed. For regular algebra this is only noticed in the intermediate 
statistics of a module (not in the answer of course), but here it becomes a 
real effect.

A good demonstration of the statistical nature is when we run the same 
program on 24 workers:
\begin{verbatim}
   ~~~CPU time till now: 1.07 sec.
   *** STATS: original  12044P 106580M 11379A : 142711
   *** STATS: optimized 41P 11745M 8359A : 20210
   ~~~CPU time till now: 1.65 sec.
   *** STATS: original  12044P 106580M 11379A : 142711
   *** STATS: optimized 42P 8873M 7417A : 16398
   ~~~CPU time till now: 7.39 sec.
   *** STATS: original  12044P 106580M 11379A : 142711
   *** STATS: optimized 23P 6331M 6003A : 12388
   ~~~CPU time till now: 376.19 sec.
     0.32 sec + 376.25 sec: 376.58 sec out of 23.07 sec
\end{verbatim}
Suddenly the final answer has 12388 operations. We also see that the total 
CPU time has increased enormously. As far as we can tell this is 
bus-congestion. It also shows that running with very many workers does not 
always give correspondingly better execution times.

\subsection{Physics examples}

For the purpose of `realistic' testing we have taken three formulas that 
were generated by the GRACE~\cite{grace,graceform} system which generates 
matrix elements for the product of one loop graphs and tree graphs. Each 
formula represents a diagram and the whole is written in terms of Feynman 
parameters. The coefficient of each combination of Feynman parameters is to 
be evaluated numerically and put in an array. A separate routine will then 
compute the corresponding integrals and multiply them by these 
coefficients. This means that the best procedure is to optimize these 
coefficients simultaneously.

The first formula comes from the reaction $e^+ e^- \rightarrow e^+ e^- 
\gamma$ and concerns a loop diagram with a 5-point function. In total there 
are 5717 terms, containing a total of 15 different symbols, which includes 
the 4 Feynman parameters. We call this formula HEP($\sigma$). A 
straightforward evaluation takes 47424 mathematical operations. For this 
example we do not do a simultaneous optimization. We just do a global 
optimization, including the Feynman parameters, just to see how this type 
of formulas can be improved. It made it also easier to create an output for 
quick testing with compilers.

The other two formulas are called $F_{13}$ and $F_{24}$ and come from the 
reaction $e^+ e^- \rightarrow \mu^+\mu^- u \overline{u}$ and are diagrams 
with a 6-point function. The $F_{13}$ formula contains 105114 terms, 5 
Feynman parameters and 24 other variables. Its direct evaluation takes 1
068 153 mathematical operations. The $F_{24}$ formula contains 836010 
terms, 5 Feynman parameters and 31 other variables. It needs 7 722 027 
mathematical operations in its raw version. Both formulas have 56 different 
combinations of Feynman parameters.

\subsection{Comparison with other algorithms}

We have compared the current implementation with results from the 
literature and programs we had access to.

\begin{table}[!ht]
\centering
\begin{tabular}{|l|r|r|r|r|}
\hline
		 & 7-4 resultant & 7-5 resultant & 7-6 resultant & HEP($\sigma$) \\ \hline
Original &         29163 &        142711 &      587880   &         47424 \\
\form O1 &          4968 &         20210 &       71262   &          6099 \\
\form O2 &          3969 &         16398 &       55685   &          4979 \\
\form O3 &          3015 &         11171 &       36146   &          3524 \\
Maple    &          8607 &         36464 &           -   &         17889 \\
Maple tryhard &     6451 & ${\cal O}(27000)$ &       -   &          5836 \\
Mathematica &      19093 &         94287 &           -   &         38102 \\
HG + cse &          4905 &         19148 &       65770   &             - \\
Haggies(Ref.~\cite{haggies})
         &          7540 &         29125 &           -   &         13214 \\ \hline
\end{tabular}
\caption{
\label{tbl:compare}
Number of operations after optimization by various programs.
The number for the 7-5 resultant with `Maple tryhard' is taken from 
ref~\cite{leiserson}. For the 7-4 resultant they obtain 6707 operations, 
which must be due to a different way of counting. The same holds for the 
7-6 resultant as ref~\cite{leiserson} starts with 601633 operations. The 
\form O3 run used $C_p=0.07$ and $10\times 400$ tree expansions.}
\end{table}

\subsection{Compiled results}

Because the object of the code optimization is to create numerical programs 
that will be shorter and faster, we also have a look at what the compiler 
can make of the code. To do this in a fair way, we must take into account 
that the pow function in C is rather inefficient. It needs two double 
arguments and then becomes very slow. Hence we made a function that takes a 
double and an integer argument, and put it in the same file as the 
optimized code. This way the compiler can make inline code and optimize 
better. For the \form optimized code this makes hardly any difference as 
there are very few calls to the power function. For the original formula 
however it makes a difference of more than a factor 10. The results are in 
Tab. \ref{tbl:Ccode}. All times were on a 3.2 GHz Xeon laptop and were 
obtained by evaluating the function $10^{6}$ times.

\begin{table}[!ht]
\centering
\begin{tabular}{|c|c|c|c|c|}
\hline
        & Format O0 & Format O1 & Format O2 & Format O3 \\ \hline
gcc -O0 & 83.543    & 11.266    & 10.200    & 6.757     \\ \hline
gcc -O1 & 14.400    &  5.278    &  4.664    & 3.171     \\ \hline
gcc -O2 & 17.091    &  5.880    &  5.266    & 3.498     \\ \hline
gcc -O3 & 17.119    &  5.686    &  5.006    & 3.302     \\ \hline
\end{tabular}
\caption{
\label{tbl:Ccode}
Execution times for the resulting C code of the physics formula 
HEP($\sigma$) in microsec. The O3 option in \form used 
$C_p=0.8$ and 3000 tree expansions. It produced 3358 terms.}
\end{table}

We have also created the outputs in 
FORTRAN and made a similar table (\ref{tbl:Fcode}).

\begin{table}[!ht]
\centering
\begin{tabular}{|c|c|c|c|c|}
\hline
             & Format O0 & Format O1 & Format O2 & Format O3 \\ \hline
gfortran -O0 & 54.687    & 11.243    & 10.192    & 6.796     \\ \hline
gfortran -O1 & 17.320    &  5.679    &  4.997    & 3.341     \\ \hline
gfortran -O2 & 17.382    &  5.689    &  5.044    & 3.318     \\ \hline
gfortran -O3 & 17.336    &  5.676    &  5.020    & 3.318     \\ \hline
\end{tabular}
\caption{
\label{tbl:Fcode}
Execution times for the resulting FORTRAN code of the physics formula 
HEP($\sigma$) in microsec. The formulas were as in Tab. \ref{tbl:Ccode}.}
\end{table}

It is interesting to note that in all cases the O1 option of the (GNU) C 
compiler gives the fastest code. In the case of the gfortran compiler this 
effect does not exist.

Of course the optimizer of the compilers is not allowed to use certain 
optimizations that we can use. A compiler is not allowed to assume that 
addition is associative in order to not upset programs that have been coded 
carefully to avoid numerical instabilities or overflow problems.

On the whole there is not much difference between the C and the FORTRAN 
code, provided we manage to have the (low) powers made inline in the C 
code. FORTRAN does this automatically.

Finally we present the corresponding table for the 7-6 resultant. This is a 
much bigger expression and we include the running times of both \form and 
the GNU C compiler.
\begin{table}[!ht]
\centering
\begin{tabular}{|r|r|r|r|r|}
\hline
             & Format O0 & Format O1 & Format O2 & Format O3 \\ \hline
  Operations &  587880   &  71262    &  55685    & 36146     \\ \hline
  \form time &  0.12     &  1.66     &   65.43   &  2398     \\ \hline
gcc -O0 time &  29.02    &  6.33     &   5.64    &  3.36     \\
         run &  119.66   &  13.61    &  12.24    &  7.52     \\ \hline
gcc -O1 time & 5098.8    & 295.96    & 199.47    & 92.09     \\
         run &  26.98    &   6.88    &   6.12    &  3.80     \\ \hline
gcc -O2 time & 4891.5    & 247.60    & 163.79    & 74.15     \\
         run &  21.87    &   7.00    &   6.22    &  3.80     \\ \hline
gcc -O3 time & 4910.4    & 276.77    & 179.24    & 79.11     \\
         run &  21.89    &   6.95    &   6.19    &  3.84     \\ \hline

gcc -O1 time & 3018.6    & 295.96    & 199.47    & 80.82     \\
         run &  24.30    &   6.88    &   6.12    &  3.58     \\ \hline
gcc -O2 time & 3104.4    & 247.60    & 163.79    & 65.21     \\
         run &  21.09    &   7.00    &   6.22    &  3.93     \\ \hline
gcc -O3 time & 3125.4    & 276.77    & 179.24    & 71.02     \\
         run &  21.02    &   6.95    &   6.19    &  3.93     \\ \hline
\end{tabular}
\caption{
\label{tbl:Cresultant}
\form run time, compilation times and the time to evaluate the compiled 
formula $10^5$ times (run). All times are in seconds. The O3 option in 
\form used $C_p=0.07$ and $10\times 400$ tree expansions. The top 
compilations are with the powers taken as macro's and the last three 
`lines' with the powers as inline functions.}
\end{table}
We can see from the table that an optimal sum-time depends on the number of 
formula evaluations\footnote{For this article we ignore the fact that the 
fastest evaluation would be to compute the original $13\times 13$ 
determinant which, considering the zeroes, would take just a few hundred 
operations.}. In all cases however one of the \form optimizations will give 
the best result. It should be noted that for this formula the O1 option of 
the compiler does not always produce the fastest code.


\section{Effects of the parameters}
\label{sect::effects}

\subsection{Scatter plots}

In this section we will study a few expressions and how the various 
parameters can influence the results of the optimization process. Because 
the MCTS process involves random numbers, the final answer is not always 
the same. We show this by means of scatter plots. In order to create scatter 
plots we have implemented an extra random number generator. \form has 
already the regular function \verb|random_|, but this function generates 
random integers. For the scatter plots we needed random values for the 
parameter $C_p$ in Eq.~(\ref{eqn::UCT}). These should be floating point 
numbers and follow a distribution. Unfortunately, currently \form does not 
have floating point numbers and the algebraic expressions do not tolerate 
them. Hence the only way we can carry floating point numbers around is as 
strings. This dictates that the preprocessor should handle them for now. 
Therefore we have created a preprocessor variable \verb|random_| which 
takes three arguments as in
\begin{verbatim}
   #define R "`random_(log,0.01,10.0)'"
\end{verbatim}
which stores a floating point value in the preprocessor variable R. The 
generation is with a logarithmic distribution and the number will be 
between 0.01 and 10.0. Internally it uses of course the same random number 
generator as the \verb|random_| function. Hence it is affected in the same 
way by the \verb|#setrandom| instruction.

Currently the \verb|random_| preprocessor variable can generate
according to two distributions: logarithmic (log) and linear
(lin). There might be more in the future if there is a demand for it. 
At the moment the only place where they can be used is in the
\verb|Format Optimize| statement for the \verb|MCTSConstant| option. People 
who would like to install their own distributions should look at the 
function PreRandom in the source file reken.c.

\subsection{Effects of MCTSconstant, MCTSnumexpand and hornerdirection}

In the first plots we show the result for HEP($\sigma$), which is the same 
expression as in section 4.4, generated with the GRACE system.
We take a random value for the constant $C_p$ and run the 
optimization for a given number of points. The final number of operations 
in the output determines the `answer' and it constitues one dot in the 
scatter plot. The program looks like
\begin{verbatim}
   Symbol amel2 zk xcp3 xcp1 x1 x5 x4 xcp2 x3 
               e2e1 e3e2 e3e1 e4e2 e4e1 EFUN;
   Off Statistics;
   .global
   #include- ReadSigma.h
   .store
   #setrandom 1021
   #do i = 1,4000
     #redefine R "`random_(log,0.01,10.0)'"
     #message mctsconstant = `R'
     Format O3,mctsconstant=`R',
            hornerdirection=backward,
            method=cse,mctsnumexpand=3000,stats=on;
     L       FF = Sigma;
     .sort
     #Optimize FF
     .store
   #enddo
   .end
\end{verbatim}
The file ReadSigma.h contains the expression and is more than 5000 lines 
long. Running the program with \tform can speed it up considerably. 
Alternatively one could run various instances of the program, each with a 
different initialization of the random number generator and with a smaller 
number of iterations in the loop. The output would look like
\begin{verbatim}
   ~~~mctsconstant = 1.482771
   *** STATS: original  1270P 39168M 5716A : 47424
   *** STATS: optimized 2P 1995M 2123A : 4122
   ~~~mctsconstant = 0.073660
   *** STATS: original  1270P 39168M 5716A : 47424
   *** STATS: optimized 2P 1916M 2287A : 4207
   ~~~mctsconstant = 0.135939
   *** STATS: original  1270P 39168M 5716A : 47424
   *** STATS: optimized 3P 1893M 2198A : 4097
                      .
                      .
                      .
\end{verbatim}
and so on.

\begin{figure}[!ht]
\includegraphics[width=\columnwidth]{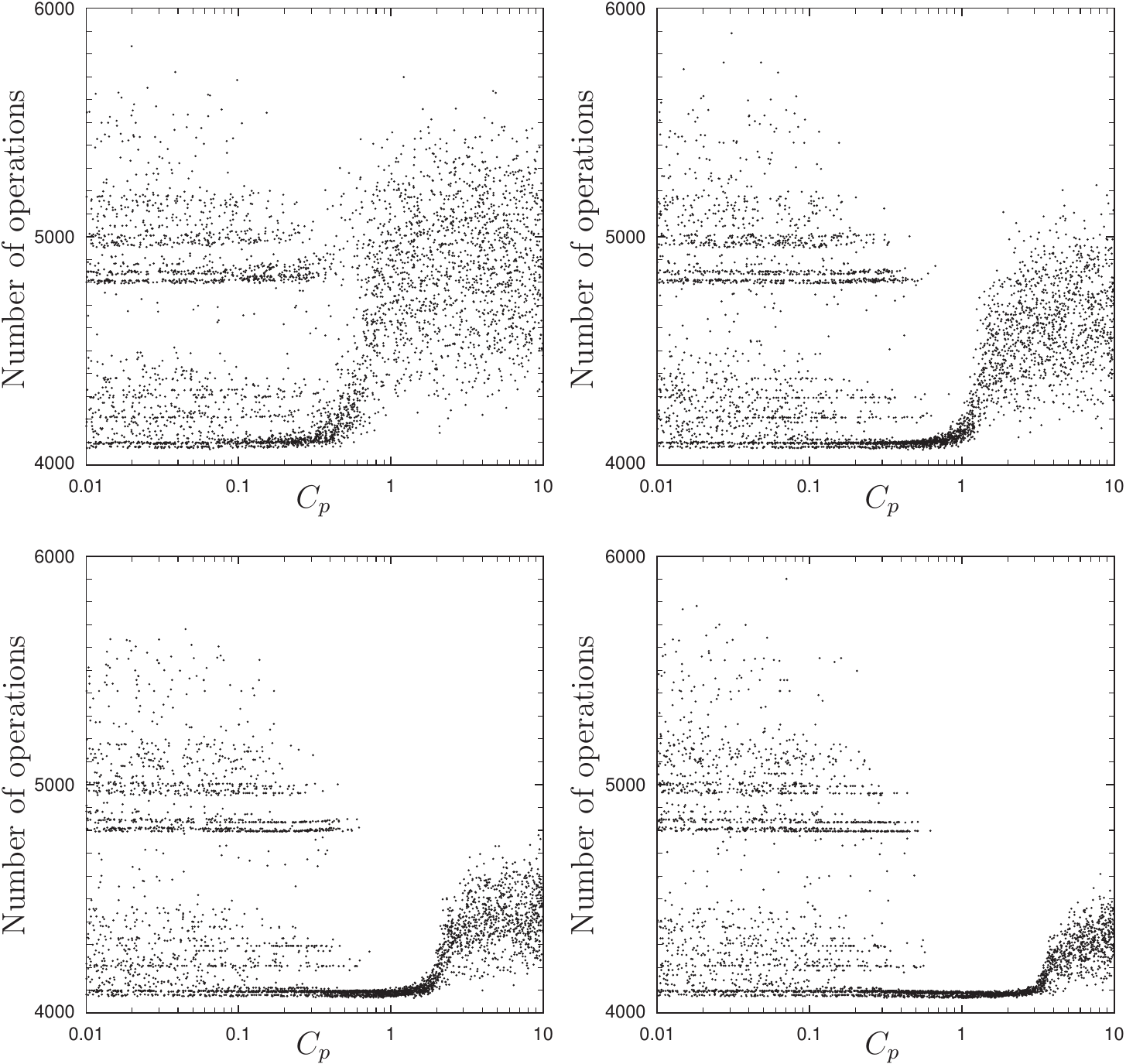}
\caption{Scatter plots for 300, 1000, 3000, 10000 points per MCTS run. Each 
plot has 4000 runs. The formula optimized is one obtained from a physics 
program (HEP($\sigma$)) as explained in the text.}
\label{fig::sigma1}
\end{figure}

Let us first look at what happens when we vary the number of tree 
expansions. In Fig.~\ref{fig::sigma1} we see scatter plots for 4 different 
values: 300, 1000, 3000 and 10000 expansions.

At the right side (larger values of $C_p$) of the plots we see a rather 
diffuse distribution. When $C_p$ is large, exploration is dominant, which 
means that at each time we try a random (new) branch and knowledge about 
the quality of previously visited branches is more or less ignored. On the 
left side there is quite some structure. Here we give a large weight to 
exploitation: we prefer to go to the previously visited branches with the 
best results. Branches that previously had a poor result may never be 
visited again. This means that there is a big chance that we end up in a 
local minimum. The plots show indeed several of those (the horizontal 
bands). When there is a decent balance between exploration and exploitation 
it becomes very likely that the program will find a good minimum. The more 
points we use the better the chance that we hit a branch that is good 
enough so that the weight of exploitation will be big enough to have the 
program return there. Hence we see that for more points the value of $C_p$ 
can become bigger. We see also that at the right side of the plots using 
more evaluations gives a better smallest value. This is to be expected on 
the basis of statistics. In the limit that we ask for more evaluations than 
there are leafs in the tree we would obtain the best value.

Clearly the optimum is that we tune the value of $C_p$ in such a way that 
for a minimum number of expansions we are still almost guaranteed to obtain 
the best result. This depends however very much on the problem. In the case 
of the formula of Fig.~\ref{fig::sigma1} this would be $C_p = 0.7$.

\begin{figure}[!ht]
\includegraphics[width=\columnwidth]{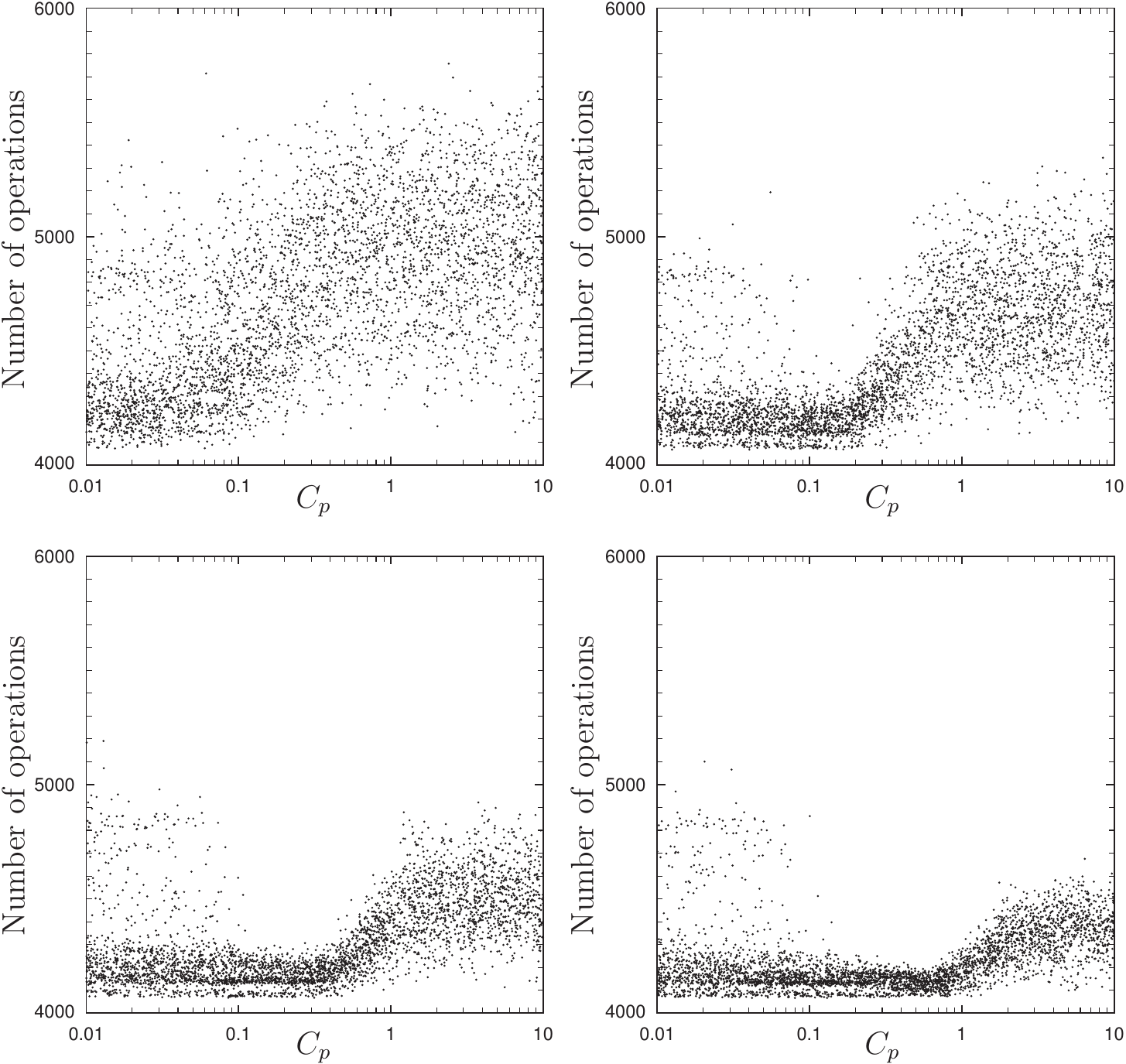}
\caption{Scatter plots for 300, 1000, 3000, 10000 points per MCTS run. Each 
plot has 4000 runs. The formula optimized is HEP($\sigma$) 
as explained in the text. Here we use the value `forward' for the 
hornerdirection parameter.}
\label{fig::sigma2}
\end{figure}

The HEP($\sigma$) formula gives the best results when we use backward as 
the value for the hornerdirection parameter. This is what we used for 
Fig.~\ref{fig::sigma1}. In the case we use forward for this parameter we 
obtain Fig.~\ref{fig::sigma2}.
Comparing the two figures shows that the first has the better potential to 
give an optimal result. It reflects the fact that the backward option gives 
a better yield of common subexpressions. This can depend very much of the 
problem as we will see.

\begin{figure}[!ht]
\includegraphics[width=\columnwidth]{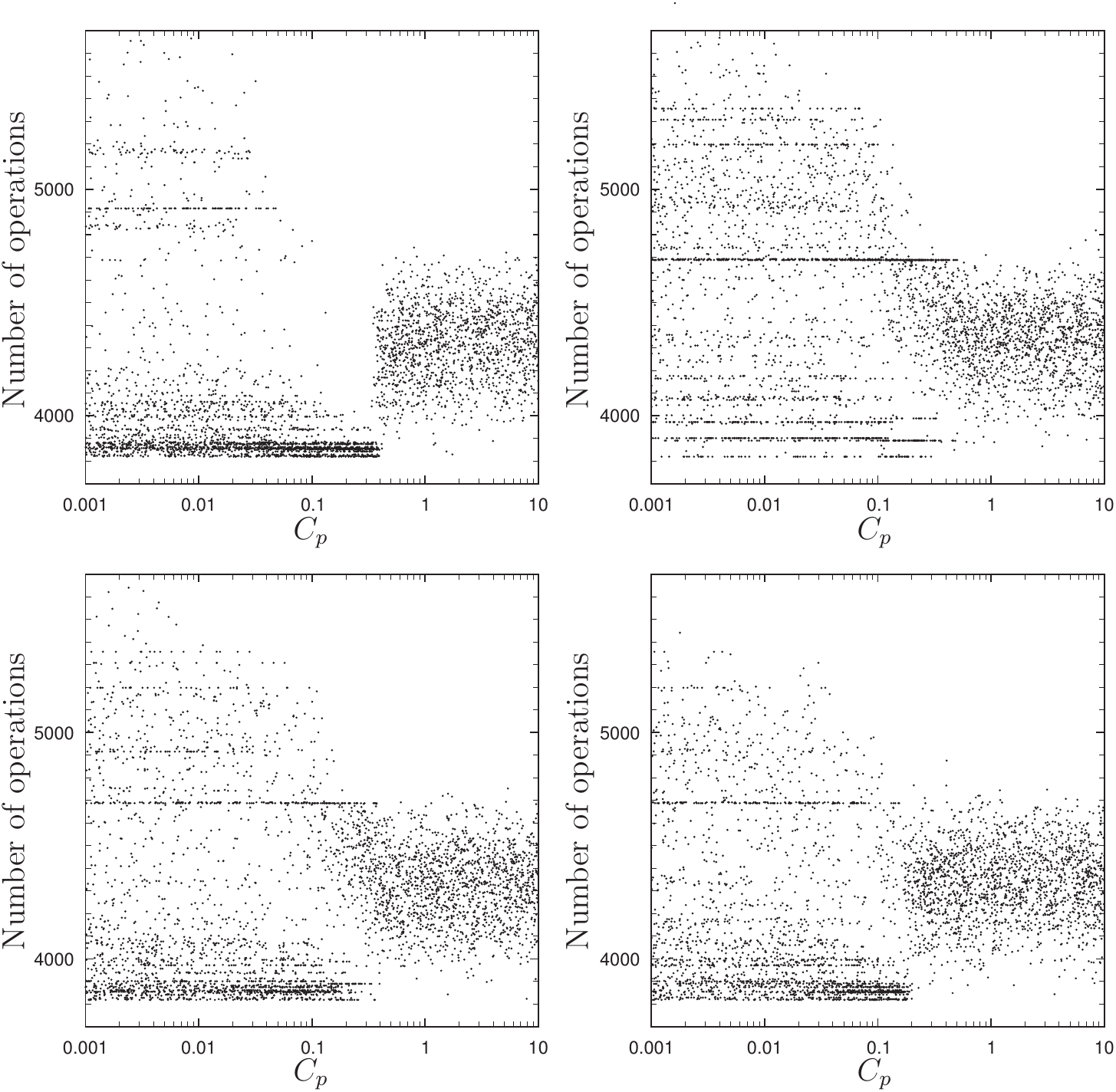}
\caption{Scatter plots for 5000 points per MCTS run. Each plot has 4000 
runs. We optimize the 7-4 resultant as explained in the text. The top left 
plot is with the forward direction and the top right plot is with the 
backward direction for the horner scheme. The bottom left plot is for the 
forwardorbackward direction and the bottom right plot is for the 
forwardandbackward direction.}
\label{fig::sigma3}
\end{figure}

With the resultants we have the opposite effect as shown in 
Fig.~\ref{fig::sigma3}. Here the backward direction is far from ideal and 
the forward direction works well, provided we go through the tree a 
sufficient number of times.
 
\begin{figure}[!ht]
\includegraphics[width=\columnwidth]{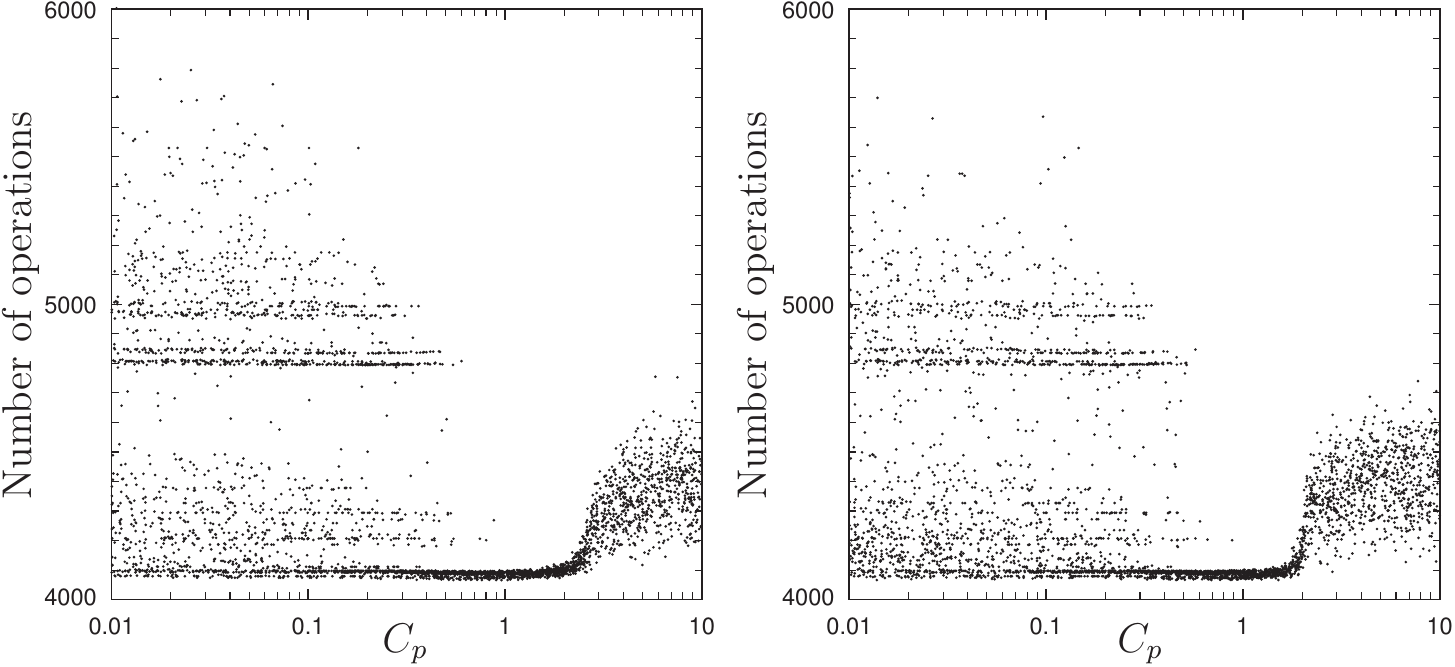}
\caption{Scatter plots for 5000 points per MCTS run. Each 
plot has 4000 runs. The formula optimized is HEP($\sigma$)
as explained in the text. The left plot is with the backwards 
direction and the right plot is with the forwardorbackward option 
(default).}
\label{fig::sigma4}
\end{figure}

The effect of the forward/backward selection shows the importance of 
selecting the proper tree structure for the problem. The more the good 
leaves are clustered on few branches, the better the MCTS can work. Because 
of the big difference between the forward/backward selections the normal 
reaction is to set the defaults to trying both orderings. In the case of 
Fig.~\ref{fig::sigma4} for the HEP($\sigma$) formula this obtains close to 
optimal results. Of course a percentage of points will be wasted by using 
the wrong direction and hence we see the random r.h.s. in the graphs move a 
little bit to the left. In Fig.~\ref{fig::sigma3} this is unfortunately 
counterproductive. This shows that once one knows the best direction for 
the problem at hand it is better to specify it. This will give a better 
efficiency. As the plots show a number of runs with a small number of 
points and relatively few tree expansions should indicate what is the best 
direction.

\subsection{Effect of MCTSnumrepeat}

\begin{figure}[!ht]
\includegraphics[width=\columnwidth]{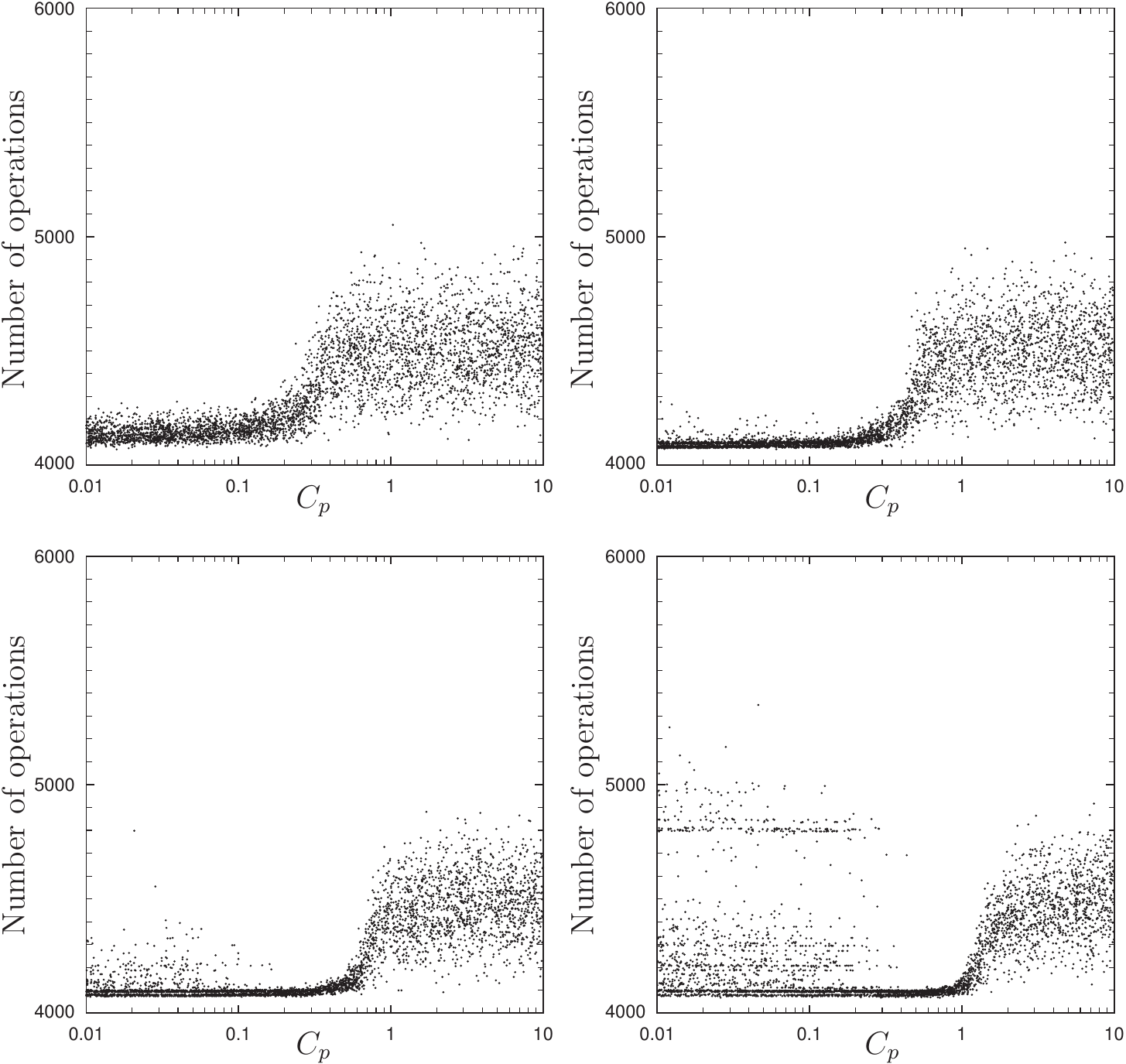}
\caption{The physics formula HEP($\sigma$) with 30 runs of 100 
expansions, 18 runs of 167 expansions, 10 runs of 300 expansions and 3 runs 
of 1000 expansions respectively. The plot with one run of 3000 expansions 
can be found in Fig.~\ref{fig::sigma1}, left bottom.}
\label{fig::sigma6}
\end{figure}

If we take another look at Fig.~\ref{fig::sigma1} we notice that in the 
left sides the distributions are nearly identical, independent of the 
number of tree expansions. This suggests a new approach: if, instead of 
3000 expansions in a single run, we take, say, 3 times 1000 expansions and 
take the best result of those, the left side of the graphs should become 
far more favorable. This is illustrated in Fig.~\ref{fig::sigma6}.
We notice a number of things here: when each run has too few points, we do 
not find a good local minimum and in the limit of runs of a single point 
per run the results revert to that of the almost random branches for large 
values of $C_p$. The multiple runs make us loose the beautiful minimum near 
$C_p=0.7$, because we do not have a correllated search of the tree. If, 
however,we have no idea what would be a good value for $C_p$ it seems best 
to select a value that is small and make multiple runs as used here, 
provided that the number of expansions is big enough for finding a decent 
local minimum in a branch of the tree. At this point it should be remarked 
that \tform and \parform will give a result that is statistically a 
little bit inferior to a run with sequential \form and the same number of 
tree expansions. In the case of sequential \form each tree expansion takes 
all previous tree expansions into account, while in the cases of \tform and 
\parform a new tree expansion has no access to the expansions that are 
still in progress inside other workers. This is particularly relevant when 
one uses many runs each with a small number of points.
 
\begin{figure}[!ht]
\centering
\includegraphics[width=440pt]{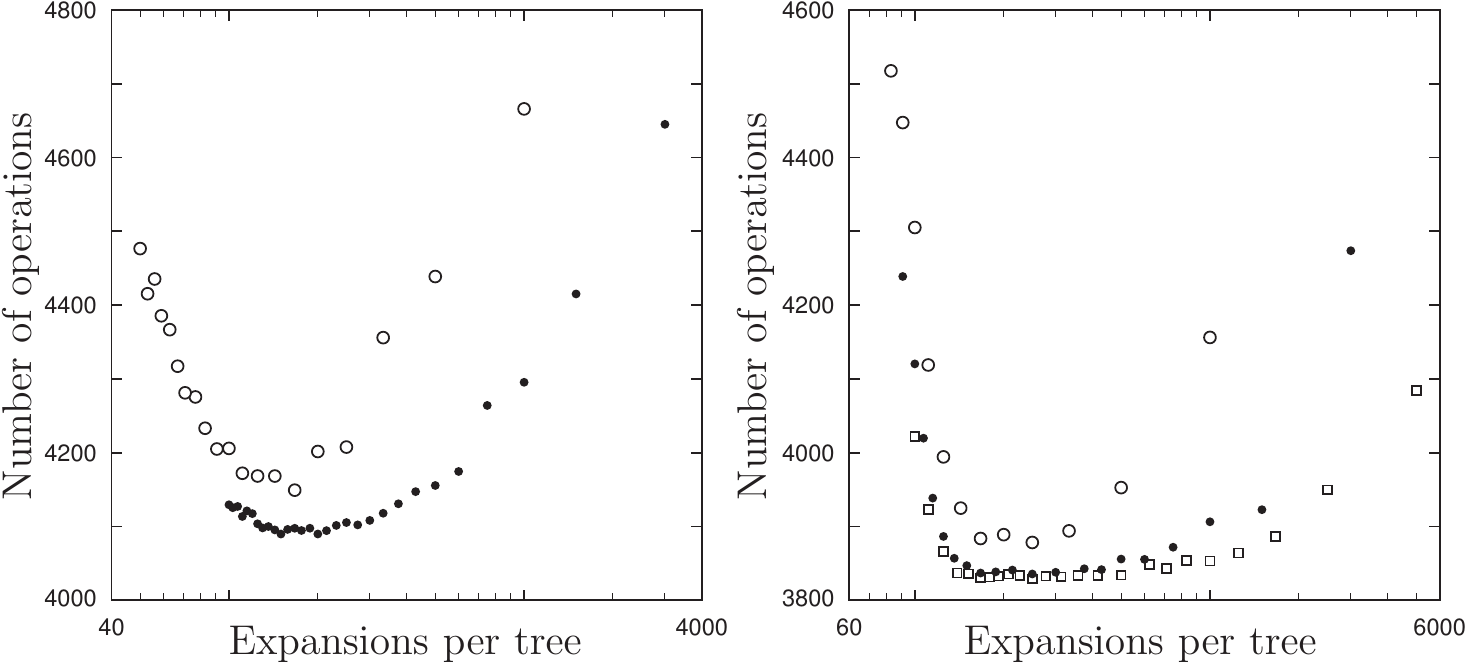}
\caption{Number of operations as a function of the number of tree 
expansions per tree. The product of the number of expansions and the number 
of runs is kept constant (1000 for the open dots, 3000 for the filled dots 
and 5000 for the open squares). The dots are average results obtained by 
running the program 50 times. The left graph is for the HEP($\sigma$) 
formula and the right graph is for the 7-4 resultant.}
\label{fig::figm}
\end{figure}

The next question is: "What is a good value for the number of tree 
expansions per run?" We investigate this in Fig.~\ref{fig::figm}. We select 
a small value for $C_p$ (0.01) and run for several values of the total 
number of tree expansions. The calculations in the left graph are for the 
formula HEP($\sigma$) and in the right graph for the 7-4 resultant. The 
minima for HEP($\sigma$) coincide more or less around 165 expansions per tree
and for the 7-4 resultant around 200 expansions per tree. We believe this 
to be correlated with the square of the number of variables. To saturate 
the nodes around a single path takes $\frac{1}{2}n(n+1)$ expansions. The 
remaining expansions are used to search around this path and are apparently 
enough to find a local minimum. The right top plot of 
Fig.~\ref{fig::sigma6} was selected with 18 trees of 167 expansions per 
tree with this minimum in mind. For this formula this seems to be the 
optimum if one does not know about the value $C_p=0.7$ or if one cannot run 
with a sufficient number of expansions to make use of its properties.

We have also made a few runs for the 7-5 and 7-6 resultants and find minima 
around 250 and 300 respectively. This suggests that if the number of 
variables is in the range of 13 to 15 a good value for the number of 
expansions is 200-250 and this will then be multiplied by the value of 
MCTSNumRepeat to obtain a good total number of tree expansions.

Similar studies of other physics formulas with more variables (${\cal 
O}(30)$) show larger optimal values for the number of expansions per run 
and less pronounced local minima. Yet also here many smaller runs can give 
better results than a single big run, provided that the runs have more than 
a given minimum number of tree expansions.

\begin{figure}[!ht]
\includegraphics[width=\columnwidth]{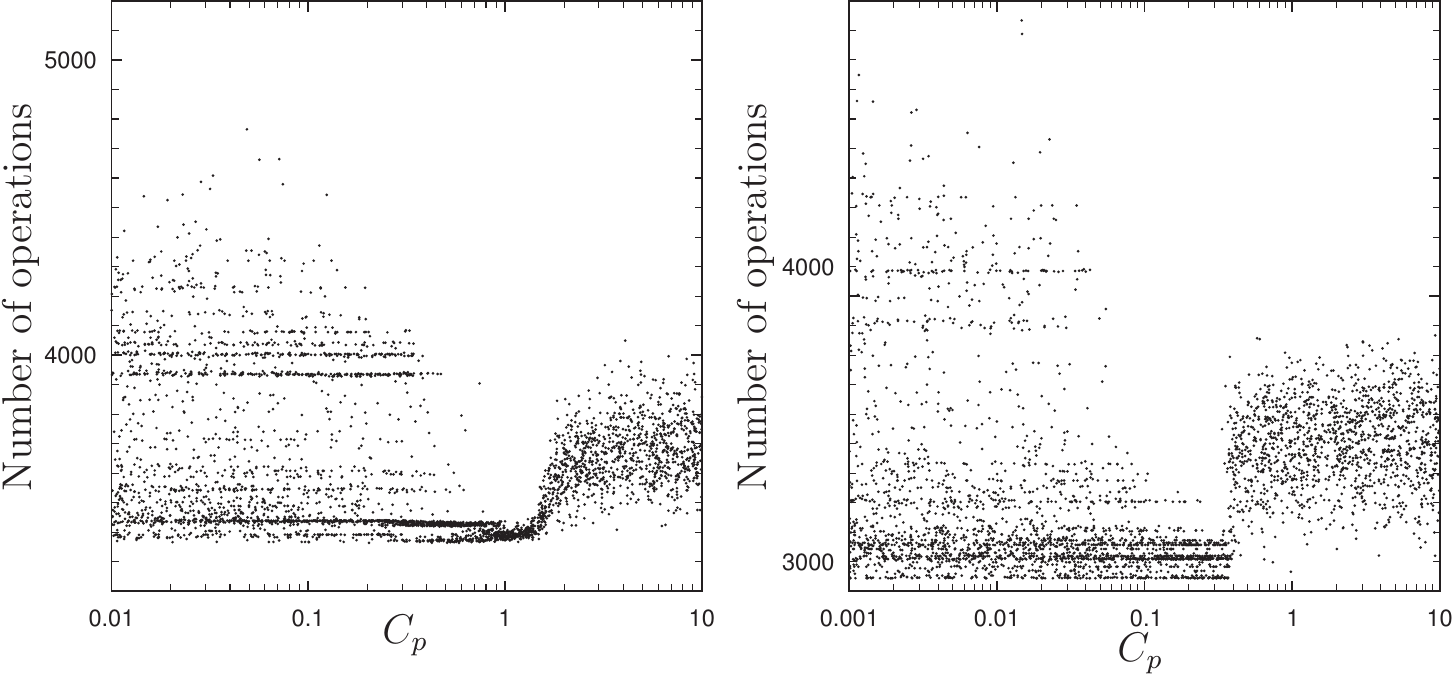}
\caption{Each plot has 4000 runs. The runs include the greedy optimization. 
The left plot is the physics formula HEP($\sigma$) with the backward 
direction and 3000 expansion per MCTS run, corresponding to 
Fig.~\ref{fig::sigma1} left bottom, and the right plot is the 7-4 resultant 
with the forward direction and 5000 expansion per MCTS run, corresponding 
to Fig.~\ref{fig::sigma3}.}
\label{fig::sigma7}
\end{figure}

In the above examples we have intentionally omitted the greedy 
optimizations because they slow the programs down and do not make 
qualitative changes in what we wanted to demonstrate. This can be seen in 
Fig.~\ref{fig::sigma7} in which we show typical results with the greedy 
optimizations included.

The plots show clearly the same behaviour as the plots without the extra 
greedy optimizations. The formulas are just 15-25\% shorter.

%
%

\section{Further improvements}

As we have mentioned before, it may pay to spend some attention to the 
formulas before sending them to the \form optimization. One generic 
improvement one may think about is a shift of variables. This could work in 
a large number of cases. Such a shift would be one of the types
\begin{eqnarray}
	x_i & \rightarrow & x_i + a x_j \\
	x_i & \rightarrow & x_i + a.
\end{eqnarray}
With such a shift the number of variables remains the same and hence the 
work for the optimizer (the size of the tree) does not become more 
complicated. This work can in principle be done in a generic procedure of 
about 100 lines. We have opted for a more sophisticated method that takes 
into account that there are various types of variables that only mix amoung 
each other. This would be the case when variables have different 
dimensions. The procedures are included with the other files (doshift.hh) 
that contain the examples of this paper. The procedures look for potential 
replacements and if they make the expression shorter (= fewer terms) the 
new expression replaces the old one. It keeps doing this until all 
potential improvements are exhausted. In principle one could do this until 
a given number of attempts has been made (the algorithm is basically 
quadratic in the number of terms) because the final improvements are 
usually rather small.

The shift procedure does not work for the resultants. The reason is that it 
looks for combinations like $\cdots(a_i x_i + a_j x_j + \cdots)$ with $x_i$ 
and $x_j$ not occurring outside the brackets, and then it tries the 
substitution $x_i \rightarrow x_i - a_j x_j / a_j$. Such combinations do 
not occur in the resultants. On the other hand in the physics formulas the 
results are very favorable. It is however important to notice that the 
order of declaration of the variables is important and hence one may have 
to experiment a little bit with this. For the three formulas we treat here 
some general rules could be set up.

The Feynman parameters were treated in a special way. In the end we want to 
bracket in them. Hence at first one might think that we should not be 
shifting them. If however we shift them too (but just combining Feynman 
parameters with each other or numbers) we obtain much better results. At 
the end we have to add some code to `unshift' the Feynman parameters, so 
that in the original parameters the code becomes a bit lengthier again. Also 
this can be optimized, although it is a bit rough on the MCTS method 
because now there are many different variables (the coefficients of the 
shifted brackets). In the Tab.~\ref{tbl:Shifted} we present the 
full results. If there are two numbers the numbers refer to the number of 
operations after the optimization plus the number of operations in the 
shift. When there are three numbers they represent the number of operations 
after the main optimization, the number of operations in the unshifting of 
the Feynman parameters and the number of operations in the shifts 
themselves.

\begin{table}[!ht]
\centering
\begin{tabular}{|c|c|c|c|c|}
\hline
             & HEP($\sigma$) & HEP($\sigma$) & $F_{13}$         & $F_{24}$         \\ \hline
Variables    &    15         &   4+11        &  5+24            &  5+31            \\
Expressions  &     1         &    35         &    56            &    56            \\ \hline
Terms        &   5 717       &  5 717        &   105 114        &   836 010        \\
Format O0    &  47 424       & 33 798        &   812 645        & 5 753 030        \\
Format O1    &   6 099       &  5 615        &    71 989        &   391 663        \\
Format O2    &   4 979       &  4 599        &    46 483        &   233 445        \\
Format O3    &   3 423       &  3 380        &    41 666        &   195 691        \\ \hline
Terms shifted&     754       &    754        &    16 439        &    78 005        \\
Format O0    &   6 278+29    &  4 402+731+15 &   123 415+605+48 &   536 127+476+57 \\
Format O1    &   1 549+29    &  1 481+409+15 &    23 459+453+48 &    68 093+336+57 \\
Format O2    &   1 216+29    &  1 146+261+15 &    17 620+330+48 &    53 131+229+57 \\
Format O3    &     976+29    &  1 012+261+15 &    13 206+322+48 &    47 379+235+57 \\ \hline
\end{tabular}
\caption{
\label{tbl:Shifted}
Results for the physics formulas in the original and the shifted versions. 
The counting of the number of operations takes the brackets in the Feynman 
parameters into account unlike ref~\cite{mctshorner}. When the number of 
variables is presented as a sum, the first number is the number of Feynman 
parameters and we do simultaneous optimizations.
}
\end{table}

The improvements are significant and in addition the \form optimization 
runs become faster because their input expressions are shorter. Also the 
compile times go down considerably.
The reason that this shifting works so well is that effectively it selects 
simple popular common subexpressions and expresses the formula in terms of 
these. This makes that the Horner scheme will be affected and so 
will the optimizations after it.

Of course other types of problems may benefit from other approaches. If one 
is resigned to not try the O3 option, the number of variables may be less 
of an obstacle and one may try 'partial replacements'. These act on only a 
fraction of the terms. Such replacements cause the number of variables to 
increase. The structure of the formulas will dictate what to try and how to 
try it. This is what we call domain specific and it falls outside the scope 
of this paper.


\section{Conclusions and outlook}
\label{sect::outlook}

The inclusion of adaptable Horner schemes, the common subexpressions and 
the greedy optimizations add considerable power to the flexibility of the 
\form output and make it competitive, if not better, than all systems we 
could compare it with. The additional MCTS methods improve the 
optimizations even further, although this goes at the cost of computer 
resources. Because the method is completely new for this field, it is not 
yet known how the various parameters should be tuned to the formulas that 
are optimized. The section on the selection of the parameters should aid 
the user in determining experimentally what is optimal for the problems at 
hand. If the computer time invested in these experiments is relatively 
unimportant, the payoff can be large. In particular the tuning of the UCT 
parameter $C_p$ and the selection of the direction of the tree search 
through the relevant variables are important and can make much difference 
in the final result. The use of several trees each with a smaller number of 
expansions can also have a great influence. We hope to be able to implement 
ways to automatically determine these parameters in future versions of 
\form.

Additional methods that would for instance recognize subexpressions that 
are powers of simpler composite objects are much harder to implement. 
Currently they have not been developed to a level that allows 
implementation. Therefore we encourage users to consider these `domain 
specific' and look for such simplifications in their programs. Often 
knowledge about the problem allows one to find a number of them.

The examples we used can be found in the \form site 
(http://www.nikhef.nl/$\sim$form) \vspace{3mm}

The work of JK and JV is part of the research program of the ``Stichting 
voor Fundamenteel Onderzoek der Materie (FOM)'', which is financially 
supported by the ``Nederlandse organisatie voor Wetenschappelijke Onderzoek 
(NWO)''.
The work of TU was supported by the DFG through SFB/TR 9
``Computational Particle Physics''.



\end{document}